\documentstyle[pre,aps,multicol,eqsecnum,epsf,rotate]{revtex}
\begin{document}
\draft
\tighten

%%%%%%%%%% Title and authors %%%%%%%%%%%%%%%%%%%%%%%%%%%%%%%%%%%%%%%%%%%%%%%%%%

\title
{\LARGE Mode coupling and renormalization group results \\
        for the noisy Burgers equation}
\author{Erwin Frey}
\address{Institut f\"ur Theoretische Physik,
         Physik--Department der Technischen Universit\"at M\"unchen, \\
         James--Franck--Stra\ss e, D--85747 Garching, Germany}
\author{Uwe Claus T\"auber $^*$}
\address{Lyman Laboratory of Physics, Harvard University, Cambridge, MA 02138; 
         and \\
         Institut f\"ur Theoretische Physik,
         Physik--Department der Technischen Universit\"at M\"unchen, \\
         James--Franck--Stra\ss e, D--85747 Garching, Germany}
\author{Terence Hwa}
\address{Physics Department, University of California at San Diego, \\ 
         3500 Gilman Drive, La Jolla, CA 92093--0319}
%\receipt{}
\date{\today}
\maketitle

%%%%%%%%%% Abstract %%%%%%%%%%%%%%%%%%%%%%%%%%%%%%%%%%%%%%%%%%%%%%%%%%%%%%%%%%%

\begin{abstract}
  We investigate the noisy Burgers equation (Kardar--Parisi--Zhang equation in
  1+1 dimensions) using the dynamical renormalization group (to two--loop
  order) and mode--coupling techniques. The roughness and dynamic exponent are
  fixed by Galilean invariance and a fluctuation--dissipation theorem. The fact
  that there are no singular two--loop contributions to the two--point vertex
  functions supports the mode--coupling approach, which can be understood as a
  self--consistent one--loop theory where vertex corrections are neglected.
  Therefore, the numerical solution of the mode coupling equations yields very
  accurate results for the scaling functions. In addition, finite--size effects
  can be studied. Furthermore, the results from exact Ward identities, as well
  as from second--order perturbation theory permit the quantitative evaluation
  of the vertex corrections, and thus provide a quantitative test for the
  mode--coupling approach. It is found that the vertex corrections themselves
  are of the order one. Surprisingly, however, their effect on the correlation
  function is substantially smaller.
\end{abstract}


%%%%%%%%%% Introduction %%%%%%%%%%%%%%%%%%%%%%%%%%%%%%%%%%%%%%%%%%%%%%%%%%%%%%%

\begin{multicols}{2}

\section{Introduction}
\label{introd}

The Kardar--Parisi--Zhang (KPZ) equation represents one of the most prominent
models describing nontrivial nonequilibrium dynamics~\cite{kpz86}. This model
equation constitutes one of the most thoroughly studied continuum theories of
kinetic roughening. It describes the height fluctuations $h({\bf x},t)$ of a
stochastically grown $d$--dimensional interface with a growth rate
$v(\bbox{\nabla} h) = \lambda (\bbox{\nabla} h)^2 / 2$ depending nonlinearly on
the local orientation of the surface,
\begin{equation}
  {\partial h \over \partial t} =
  \nu \bbox{\nabla}^2 h + {\lambda \over 2} (\bbox{\nabla} h)^2 + 
  \eta({\bf x},t) \ .
\label{kpz}
\end{equation}
The ($\nu \bbox{\nabla}^2 h$)--term mimics a surface tension, and acts to
smooth the interface, while the uncorrelated Langevin noise $\eta({\bf x},t)$
tends to roughen the interface and entails the stochastic nature of any growth
process.  Its first moment vanishes, and its second moment is given by
\begin{equation}
  \langle \eta({\bf x},t) \eta({\bf x}',t') \rangle =
  2 D \delta^{(d)}({\bf x}-{\bf x}') \delta(t-t') \ ;
\label{2nd_moment_noise}
\end{equation}
note that in general the coefficients $\nu$ and $D$ are not related in any
simple manner, in contrast to near--equilibrium situations where Einstein
relations connect damping constants and noise correlations.

\noindent{\em Dynamic scaling.} 
The interface fluctuations are characteristically scale--invariant, i.e., the
height profile obtained by a {\em self--affine rescaling} $h'({\bf x},t) =
b^{-\chi} h(b {\bf x}, b^z t)$ is, in a statistical sense, equivalent to
$h({\bf x},t)$. As a consequence, for sufficiently large ${\bf x}_0$ and $t_0$,
such that the process is already beyond the initial transient region, the
correlation function
\begin{equation}
  C({\bf x},t) = \langle 
  [ h({\bf x}+{\bf x}_0,t+t_0) - h({\bf x}_0,t_0) ]^2 \rangle
\end{equation}
obeys the generalized homogeneity relation ($x = |{\bf x}|$)
\begin{equation}
  C(x,t) = b^{-2 \chi} C(b x,b^z t) \ .
\end{equation}
Upon choosing the scaling parameter $b=1/x$ we obtain the dynamic scaling form
\begin{equation}
  C(x,t) = x^{2 \chi} {\hat C}(t/x^z) \ .
\end{equation}
In the asymptotic limits $t \to 0$ and $x \to \infty$, the scaling function
${\hat C}(t/x^z)$ displays power law behavior and hence
\begin{equation}
  C(x,t) =\cases{A x^{2 \chi}  & \,  {\rm for} $t \to 0$ \ , \cr
                 B t^{2 \chi/z}& \,  {\rm for} $x \to 0$ \ .}
\end{equation} 
The transverse wandering of the interface may be characterized by a
perpendicular correlation length $\xi_\perp (x) \propto \sqrt{C(x,t=0)} \propto
x^{\chi}$ with the {\em roughness exponent} $\chi$. The temporal increase of
surface roughness is described by a parallel correlation length
$\xi_\parallel(t) \propto t^{1/z}$ with the {\em dynamic exponent} $z$.

Many growth phenomena show the above dynamic scaling of the interface
fluctuations, but with values for the critical exponents different from those
obtained for the KPZ equation. Nevertheless, the KPZ equation has become the
starting point for our understanding of nonequilibrium dynamics and strong
coupling behavior.

\noindent {\it Phenomenology of the KPZ equation.} 
The phenomenology of the KPZ equation is now well known~\cite{ft94}. Below the
{\em lower critical dimension} $d_{\rm lc} = 2$ there appear two
renormalization--group (RG) fixed points, namely an infrared(IR)--unstable
Gaussian fixed point and an IR--stable strong--coupling fixed point describing
a smooth and a rough interface, respectively. For dimensions $d > 2$ there
exists a nonequilibrium phase transition from a weak--coupling phase for small
effective coupling constants $g = \lambda^2 D / \nu^3$, where the nonlinearity
is irrelevant (in the RG sense), to a strong--coupling phase which seems to be
inaccessible through perturbative methods~\cite{ft94,lae95}. The scaling
exponents in the strong--coupling phase have been determined by numerical
methods~\cite{hz95,ahk93} and self--consistent mode--coupling
approaches~\cite{dmk94,bc93,tu94}. The results obtained from mode--coupling
theory suggest the existence of an {\em upper critical dimension}
$d_{\rm uc}=4$~\cite{mbd95}. This result is supported by functional RG
calculations~\cite{hal90,nl91} and renormalization group
arguments~\cite{lae95,tf95}. In the numerical simulations, however, the dynamic
critical exponent $z$ for the transient roughening of an initially flat
interface is found to be smaller than $z_0=2$ for all dimensions accessible to
a numerical analysis~\cite{ahk93}, i.e., there is no indication of any upper
critical dimension. This discrepancy between mode--coupling theory and
numerical results has yet to be resolved and constitutes one of the most
important issues of current theoretical research.

\noindent {\it Mapping to other models.}
The KPZ equation is closely related to a variety of other problems ranging from
fluid dynamics governed by the Burgers equation~\cite{fns77} to equilibrium
systems with quenched disorder, namely directed polymers in random
environments~\cite{dp,dp1}. Most of these mappings and relations are strictly
valid for the one--dimensional case only. In order to assist the reader with
the transfer of the results obtained in the main part of this paper to related
systems, we provide a short account of some of the most important issues.

The transformation ${\bf v} = - \bbox{\nabla} h$ leads to a Langevin equation
for a randomly stirred fluid
\begin{equation}
  {\partial {\bf v} \over \partial t} +
  \lambda \left( {\bf v} \cdot \bbox{\nabla} \right) {\bf v} =
  \nu \bbox{\nabla}^2  {\bf v} - \bbox{\nabla} \eta ( {\bf x}, t ) \ ,
\label{noisy_burgers}
\end{equation}
which in the case $\lambda = 1$ represents a $d$--dimensional generalization of
the noisy Burgers equation~\cite{fns77}. The long--time and large--distance
behavior of the Burgers equation, describing the dynamics of a vorticity--free
velocity field, and the Navier--Stokes equation, characterizing an
incompressible fluid, have been analyzed by Forster, Nelson, and Stephen in the
framework of dynamical renormalization group theory to one--loop
order~\cite{fns77}. These authors have shown that the fluctuation--dissipation
theorem, valid in $d=1$ only (see App.~\ref{app_a}), together with a Ward
identity resulting from the Galilean invariance of the fluid equation of motion
allow the determination of the dynamic critical exponent $z$ in $d=1$ to be
exactly $z = 3/2$. Their RG analysis has recently been extended to two--loop
order~\cite{sp94,ft94,tf95}.

Another model of surface roughening, which is governed by the same nonlinearity
as the KPZ equation, is the Kuramoto--Sivashinsky (KS) equation~\cite{ks77}. In
contrast to the KPZ equation the KS equation is completely deterministic
\begin{equation} 
  {\partial h \over \partial t} =
  -\nu \bbox{\nabla}^2 h - \bbox{\nabla}^4 h + 
  {\lambda \over 2} (\bbox{\nabla} h)^2 \ ,
\label{kuramoto_sivashinsky}
\end{equation}
and is characterized by a band of unstable modes at small wave vectors. (Note
that $\nu > 0$.)  Numerical simulations of the discretized one---dimensional KS
equation have recently demonstrated that the large--scale dynamical
correlations are described by the (1+1)--dimensional KPZ equation~\cite{skj92}.
A derivation of the KPZ equation from the KS equation has also been given in
Ref.~\cite{ch95}, where the effective parameters of the KPZ equation have been 
determined from the numerics of the microscopic chaotic dynamics of the KS
equation. For $d \geq 2$, however, the results~\cite{lp92} are still 
controversial.

Recently, Golubovi\'c, and Wang succeeded in mapping the equilibrium
statistical mechanics of a two--dimensional smectic--A liquid crystal onto the
nonequilibrium dynamics of the (1+1)--dimensional stochastic nonlinear KPZ
(noisy Burgers) equation~\cite{gw92}. Kashuba has shown that there exists a
one--to--one relationship between the Hamiltonian describing the nonlinear
elasticity of a two--dimensional smectic--A liquid crystal and the Hamiltonian
characterizing the long--range spin fluctuations in a two--dimensional planar
ferromagnet subject to (two--dimensional) dipolar forces~\cite{kas94}.  These
relationships thus provide an interesting, exact approach to study the
anomalous elasticity of smectic--A liquid crystals, as well as the spin
fluctuations in the ordered phase of a dipolar planar ferromagnet in two
dimensions, provided the corresponding KPZ growth model can be solved exactly,
or at least to a high degree of accuracy.

A number of somewhat more exotic relationships have been found very recently.
E.g., the kinetics of the annihilation process $A+B \to 0$ with driven
diffusion was mapped onto the (1+1)--dimensional KPZ equation~\cite{ikr95}, and
the formal equivalence of the continuum limit of the Heisenberg equation of
motion of a certain spin--1/2 chain with the Fokker--Planck equation
corresponding to the noisy Burgers equation was demonstrated~\cite{fem95}.
Besides these various mappings and relationships, which are valid in (1+1)
dimensions only, the KPZ equation is also closely related to the dynamics of a
sine--Gordon chain~\cite{ks89}, the driven--diffusion equation~\cite{dds}, and
directed paths in a random media~\cite{dp}.

\noindent {\it Invariances of the noisy Burgers equation.}
The one--dimensional KPZ equation is special in several ways. First, there is a
huge list of mappings onto related models as described above. Hence any
advances in understanding the growth model will have broad implications on many
physical problems. Second, the noisy Burgers equation has two important
``symmetry'' properties, namely Galilean invariance and detailed balance. The
Galilean invariance~\cite{fns77} of the one--dimensional hydrodynamic equation
(\ref{noisy_burgers}), corresponds to an invariance of the stochastic growth
model with respect to an infinitesimal tilt of the surface, $h \rightarrow h +
{\bf v} \cdot {\bf x}$, ${\bf x} \rightarrow {\bf x} - \lambda {\bf v} t$. As a
consequence of this symmetry, one finds that the amplitude of the nonlinearity
$\lambda$ is invariant under RG transformations, which in turn implies an
exponent identity relating the roughness exponent $\chi$ to the dynamic
exponent $z$,
\begin{equation}
  \chi + z = 2 \ .
\label{exponent_identity}
\end{equation}
Whereas the latter invariance is valid for any dimension $d$, the detailed
balance property of the KPZ equations holds in $d=1$ only (see Appendix~A). It
can be shown~\cite{kpz86} that the Fokker--Planck equation corresponding to the
(1+1)--dimensional KPZ equation has the stationary solution
\begin{equation}
  {\cal P}_{\rm st} (h) \propto \exp \left[ - {\nu \over 2 D}
  \int dx \left( {\partial h \over \partial x} \right)^2 \right] \ ;
\label{stationary_solution}
\end{equation}
this implies that the roughness exponent is $\chi = 1/2$, as if the
nonlinearity were entirely absent. Together with the exponent identity
(\ref{exponent_identity}), one thus finds for the dynamic exponent $z = 3/2$.

\noindent{\it Scaling of the (1+1)--dimensional KPZ equation.}
As a consequence of the above invariance properties of the nonlinear Langevin
equation (\ref{kpz}) one can show that the height--height correlation function
obeys the following scaling law~\cite{hf91}
\begin{equation}
   C(x,t) = A x^{2\chi} F(\lambda \sqrt{A} t/x^z) \ .
\label{invariance_scaling}
\end{equation}
The argument of the scaling function is now dimensionless, and the scaling
function itself is {\em universal}. It acquires the asymptotic form
\begin{equation}
  F(\xi) = \cases{ 1                      & for $\xi \to 0 \ ,$ \cr
                   (\xi/2g^{*})^{2\chi/z} & for $\xi \to \infty \ .$}
\end{equation}
The RG fixed point of the (1+1)--dimensional KPZ equation turns out to be a
strong--coupling fixed point. As discussed above, despite this fact the
roughness and the dynamic exponent are known {\em exactly} as a consequence of
the particular invariance properties of the one--dimensional case. The scaling
function $F(\xi)$ has been calculated using a non--perturbative mode--coupling
approach~\cite{hf91}. Striking agreement with the results of direct numerical
simulations~\cite{kmh92,t92,af92} were found.

The non--perturbative mode--coupling approach essentially consists in a
resummation of the perturbation theory, such that all propagator
renormalizations are properly taken into account, while the vertex corrections
are neglected. This is clearly a very ad--hoc and uncontrolled procedure;
nevertheless, mode--coupling theories have been remarkably successful in
applications to many areas of condensed matter theory, such as structural glass
transitions~\cite{goetze90}, critical dynamics of magnets~\cite{kk70,frey94},
binary mixtures~\cite{kk70,kk76}, and others~\cite{kk76}. In all those fields,
it has been found that mode--coupling theory is capable of describing
experiments in a quantitative manner. The factorisation approximation in the
above mode coupling concepts is also known in the theory of hydrodynamic
turbulence as Kraichnan's Direct Interaction Approximation ~\cite{my75}.

The present work is motivated by this fact, and furthermore by the striking
agreement of the mode--coupling results and those obtained from numerical
simulations for the KPZ equation. In what follows, we will try to give a
systematic analysis of the mode--coupling approach using the field--theoretic
formulation of Langevin dynamics~\cite{bjw76,dd75,msr73}. In particular, the
fact that there are no singular two--loop contributions to the two--point
vertex functions in perturbation theory in $d=1$ strongly supports the
mode--coupling approach. As the IR singularities, i.e., the exponents $z$ and
$\chi$, are exactly known, the self--consistent treatment is expected (and
found) to reproduce the scaling functions to a high degree of accuracy. In
addition, we shall analyze vertex corrections in order to understand the range
of validity of the mode--coupling approach. Our explicit results for the vertex
corrections, as obtained from (exact) Ward identities, as well as from
second--order perturbation theory allow for a quantitative estimate of the
systematic errors enshrined in the mode--coupling approach. Since this specific
type of self--consistent treatment is used in many areas of theoretical
physics, albeit under different nomenclature, we hope that this work will shed
some light on its applicability, limitations, and possible extensions.

\noindent {\it Outline.} 
The outline of the paper is as follows. In the subsequent section we summarize
results from previous RG studies, discuss their relevance for the
mode--coupling approach, and provide those explicit results which are needed in
subsequent calculations. The formulation of the mode--coupling theory is
discussed in Sec.~III, as well as the solution of the self--consistent
mode--coupling equations for the noisy Burgers equation. In addition to the
scaling functions in the thermodynamic limit, finite--size corrections are
explored. The size of the vertex corrections is estimated from the (exact) Ward
identities stemming from Galilean invariance, as well as from the explicit
two--loop perturbational contributions. In the bulk of the present work, we
shall refer solely to the (1+1)--dimensional KPZ equation; however, whenever
more general statements in $d$ dimensions are possible, this restriction to
$d=1$ is relaxed. We conclude with a brief summary and a discussion of some of
the remaining open problems.

%%%%%%%%%% Renormalization Group Theory %%%%%%%%%%%%%%%%%%%%%%%%%%%%%%%%%%%%%%%

\section{Results from renormalization group theory}
\label{rg_theory}

We start by reviewing some known results from perturbational renormalization
group theory \cite{kpz86,sp94,ft94}, specializing to 1+1 dimensions. This
section also contains the explicit expressions for the vertex corrections to
the two--point vertex functions to two--loop order. In this section, as well as
in the Appendix, unrenormalized quantities are denoted by a subscript ``$0$''.

\subsection{Dynamic functional}
\label{rg_a}

We start with a brief description of the field--theoretical formulation of
Langevin--type dynamics~\cite{bjw76,dd75}. The stochastic forces
$\eta({\bf x},t)$ obeying $\langle \eta({\bf x},t) \rangle = 0$ and
Eq.~(\ref{2nd_moment_noise}) can be taken to be Gaussian distributed,
\begin{equation}
  W[\eta] \propto \exp \left[ - {1 \over 4 D_0} \int \! d^dx \! \int \! dt
                              \eta^2 ({\bf x},t) \right] \ .
\label{weight_function}
\end{equation}
Using the equation of motion (\ref{kpz}), we can eliminate the noise term; with
an additional Gaussian transformation introducing Martin--Siggia--Rose
auxiliary fields $\tilde h$~\cite{msr73} the ensuing probability distribution
$P[h]$ for the height fluctuations may be further linearized, and thus the
original nonlinear stochastic equation of motion can be reformulated in terms
of a generating functional~\cite{ft94}
\begin{eqnarray}
  Z[j,{\tilde j}] = &&\int {\cal D}[h] {\cal D}[i {\tilde h}] 
  \exp \Biggl( {\cal J}[{\tilde h},h] \nonumber \\
  &&+ \int \! d^dx \! \int \! dt \left[ {\tilde j} {\tilde h} + j h \right] 
               \Biggr) \ ,     
\label{generating_functional}
\end{eqnarray}
with the Janssen--De Dominics functional given by
\begin{eqnarray}
  {\cal J}[{\tilde h}, h] = 
 &&\int \! d^dx \! \int \! dt \Biggl\{ D_0 \, {\tilde h}{\tilde h} \nonumber \\
 &&- {\tilde h} \left[ {\partial h \over \partial t} - \nu_0 \bbox{\nabla}^2 h 
   - {\lambda_0 \over 2} (\bbox{\nabla} h)^2 \right] \Biggr\} \ .
\label{jansen}
\end{eqnarray}
Correlation and response functions can now be expressed as functional averages
with weight $\exp \left\{ {\cal J} [{\tilde h},h] \right\}$. Upon separating
the dynamic functional into a quadratic and a nonlinear part, a standard
perturbation theory can be formulated, where the cumulants $G_{{\tilde N},N}$
of the correlation and response functions are defined by functional derivatives
of $F [{\tilde j}, j] = \ln Z[{\tilde j},j]$ with respect to the sources
${\tilde j}$ and $j$, respectively. Vertex functions $\Gamma_{{\tilde N},N}$
are then obtained from the cumulants by a Legendre transformation,
\begin{equation}
  \Gamma [{\tilde h},h] = - F  [{\tilde j},j] +
  \int d^d x \int dt ( {\tilde h} {\tilde j} + h j ) \, ,
\label{legendre}
\end{equation}
where
\begin{equation} 
  h = { \delta F / \delta j}\, , \quad {\rm and} \quad
  {\tilde h} = {\delta F /\delta {\tilde j}} \, .
\label{def_fields}
\end{equation}
We finally note that the functional determinant originating in the variable
change from the noise fields $\eta$ to the height fluctuations $h$ serves to
exactly cancel the acausal contributions to the perturbation series, thus
leaving only those Feynman diagrams with correct time ordering in the response
propagators~\cite{bjw76,ft94}.

\subsection{Two--point Vertex functions and renormalization}
\label{rg_b}

We can now proceed to study the renormalization of the KPZ equation in one
dimension. As discussed in detail in Ref.~\cite{ft94}, the Ward identity
stemming from the Galilean invariance of the Burgers equation shows that the
nonlinearity $\lambda = \lambda_0$ does not renormalize. This leaves the
renormalization of the surface tension (diffusion coefficient) $\nu_0$ and of
the noise correlation strength $D_0$, which may be inferred from studying the
two--point vertex functions $\partial_{q^2} \Gamma_{{\tilde h}h}(q,\omega)$ and
$\Gamma_{{\tilde h}{\tilde h}}(q,\omega)$, respectively; because of the
fluctuation--dissipation theorem valid (only) in $d=1$ (see Appendix A), these
coefficients are actually proportional to each other and must therefore
renormalize in the same way. In Appendix B, we list the Feynman diagrams and
the corresponding analytical expressions for $\Gamma_{{\tilde h} h}(q,\omega)$
to two--loop order (second--order perturbation theory in $\lambda$),
specializing the results of Ref.~\cite{ft94} to $d=1$. Upon collecting these
terms, splitting the vertex functions into regular and (UV) singular parts,
$\Gamma_{{\tilde h} h} = \Gamma_{{\tilde h} h}^{reg} +
\Gamma_{{\tilde h} h}^{sing}$, eventually the following comparatively simple
results are obtained:
\begin{eqnarray}
  &&{\Gamma_{{\tilde h} h}^{\rm reg}(q,\omega) \over i \omega + \nu_0 q^2} =
    - {\lambda^4 D_0^2 \over 2 \nu_0^3} q^2 \int_p \! \int_k \!
    {q_- \over i \omega + \nu_0 q_+^2 + \nu_0 q_-^2} \times \nonumber \\
  &&\times {1 \over {\tilde q}_-
    [i \omega + \nu_0 {\tilde q}_+^2 + \nu_0 {\tilde q}_-^2]
    [i \omega + \nu_0 q_+^2 + \nu_0 {\tilde q}_-^2 + \nu_0 k^2]} \label{A18} \\
  &&\Gamma_{{\tilde h} h}^{\rm sing} (q,\omega) = i \omega + \nu_0 q^2 
    \nonumber \\
  &&\qquad \qquad \qquad + {\lambda^2 D_0 \over 2 \nu_0} q^2 \int_p \!
   {1 \over i \omega + \nu_0 q_+^2 + \nu_0 q_-^2} \ ; \label{A19}
\end{eqnarray}
here we have introduced the abbreviations $q_\pm = (q/2) \pm p$, ${\tilde
  q}_\pm = q_\pm \pm k$, and $\int_p = \int_{-\infty}^{+\infty}dp/2\pi$.  Note
that the singular term stems entirely from the one--loop diagram (the
expression involving only one internal momentum p), while the (UV) singular
two--loop contribution vanishes. The second--order term in the perturbation
expansion thus yields merely regular corrections to the scaling functions. The
second relevant vertex function can be written as $\Gamma_{{\tilde h}{\tilde
    h}}(q,\omega) = - 2 D_0 {\rm Re} \Gamma_{{\tilde h}h}(q,\omega) / \nu_0
q^2$, which allows us to define the wavenumber-- and frequency--dependent
diffusion coefficient as
\begin{equation}
   \nu(q,\omega) = {1 \over \nu_0 q^2} {\rm Re} \Gamma_{{\tilde h} h}(q,\omega)
    = - {1 \over 2 D_0} \Gamma_{{\tilde h}{\tilde h}}(q,\omega) \ ,
 \label{fdt1}
\end{equation}
confirming the validity of the fluctuation--dissipation theorem of Appendix 
A~\cite{ft94,sp94}.

In evaluating those contributions which become singular as the critical
dimension $d_{\rm lc} = 2$ is approached, one has to be careful to choose a
normalization point (NP) where either $q$ or $\omega$ are finite, in order not 
to interfere with the IR singularities, which would also appear as poles in 
$\varepsilon = d-2$ (for a more detailed discussion, see Refs.~\cite{ft94} and 
\cite{tf95}). A convenient choice is NP: $q = {\bf 0}$,
$i \omega / 2 \nu = \kappa^2$; with $g_0 = \lambda^2 D_0 / \nu_0^3$ one thus
arrives at
\begin{eqnarray}
  \Gamma_{{\tilde h}{\tilde h}}(q,\omega)^{\rm sing}_{\rm NP} = - 2 D_0 \Biggl[
  1 &&+ {g_0 \over 4} \int_p {1 \over \kappa^2 Z + p^2} \Biggr] \ ,
\label{A20} \\
  {\partial \over \partial q^2}
  \Gamma_{{\tilde h} h}(q,\omega)^{\rm sing}_{\rm NP} = \nu_0 
  \Biggl[ 1 &&+ {g_0 \over 4} \int_p {1 \over \kappa^2 Z + p^2} \Biggr] \ ,
\label{A21}
\end{eqnarray}
where $Z$ is the renormalization factor for both $\nu$ and $D$. The remaining
singular integral is readily evaluated using the dimensional regularization
scheme
\begin{equation}
   \int_p {1 \over \mu^2 + p^2} = - {C_d \mu^\varepsilon \over \varepsilon} \ ,
 \label{B1} \\
\end{equation}
where $C_d = \Gamma(2-d/2) / 2^{d-1} \pi^{d/2}$ is a geometry factor, and
$C_1 = 1/2$. Note that in this evaluation at {\it fixed} dimension $d=1$, no
expansion with respect to $\varepsilon = d-2$ was applied; the latter parameter
was merely used to effectively count the singularities in the integrals that
would appear at $d_{\rm lc}=2$, when they are generalized to arbitrary
dimension $d$. These ultraviolet (UV) poles may now be absorbed in renormalized
quantities $D = Z D_0$ and $\nu = Z \nu_0$, with the renormalization constant
\begin{equation}
  Z = 1 - {g_0 \kappa^\varepsilon \over 8 \varepsilon}
        + {g_0^2 \kappa^{2 \varepsilon} \over 128 \varepsilon} \ .
\label{Z}
\end{equation}
Defining the renormalized coupling
\begin{equation}
  g = {g_0 \over Z^2 \kappa} \ , \label{fp}
\end{equation}
we can now readily calculate Wilson's flow functions,
\begin{eqnarray}
  \zeta(g) &&= \kappa \partial_\kappa \vert_0 \ln Z = - g/8 \, , \\
  \beta(g) &&= \kappa \partial_\kappa \vert_0 g = g (d-2 - 2 \zeta)
                                                = g (-1 + g/4)
\end{eqnarray}
in $d=1$. Searching for zeros of the beta function yields the (IR) stable
nontrivial fixed point
\begin{equation}
  g^* = 4 \ ,
\end{equation}
from which the critical exponents
\begin{eqnarray}
  \chi &&= -\zeta(g^*) = 1/2 \ , \label{chi} \\
     z &&= 2 + \zeta(g^*) = 3/2  \label{z}
\end{eqnarray}
can be deduced. Note that these explicit results fulfill the exponent sum rule
(\ref{exponent_identity}); of course, as these exponents can already be
determined from this identity and the additional constraint of the
fluctuation--dissipation theorem (see Appendix A), this rather serves as a
check for the calculations. Note that the remarkable cancellation of the
singular two-loop contributions has been essential here from the diagrammatic
point of view.

In Ref.~\cite{ft94}, the renormalization group approach is carried out in
arbitrary space dimension $0 \leq d < 4$. For $d > 2$ an expansion with respect
to $\varepsilon = d-2$ can be pursued, and was in fact recently carried through
to arbitrary order in the perturbation series by L\"assig \cite{lae95}. For
$d<2$, on the other hand, one may note that the fixed point coupling $g^*
\propto d$ approaches zero for $d \rightarrow 0$, and the results may be cast
into an expansion about zero space dimension \cite{tf95}.

\subsection{Two--loop scaling functions}
\label{rg_c}

For later use, we now summarize the results from the second--order perturbation
theory once more, albeit with some slight changes. First, we explicitly
separate the zero-- and one--loop contributions, and the two-loop contributions
due to propagator and vertex renormalizations. Second, we take
``self--consistent'' propagators, i.e., we generalize $\nu_0$ and $D_0$ to a
$q$--dependent quantity according to Eq.~(\ref{fdt1}), however neglecting its
frequency dependence. This is in the spirit of the Lorentzian approximation in
mode--coupling theory, to be discussed below; its formal advantage is that the
pole structure in the complex frequency plane remains unaltered, and therefore
the results from Appendix B may be readily generalized. The zero-- and
one--loop contributions to $\Gamma_{{\tilde h}h}(q,0)$ thus read (see
Fig.~\ref{2loop_diagrams}a,b):
\begin{equation}
  \Gamma_{{\tilde h}h}^{(1)}(q,0) = q^2 \left[ \nu(q) + {\lambda^2 \over 2}
  \int_p \! {1 \over \nu(q_+) q_+^2 + \nu(q_-) q_-^2} \right] \ ;
\label{1loop}
\end{equation}
similarly, the two--loop contribution due to propagator renormalization
(Fig.~\ref{2loop_diagrams}c--f) becomes
\begin{eqnarray}
  &&\Gamma_{{\tilde h}h}^{(2,{\rm p})}(q,0) = - q^2 {\lambda^4 \over 2}
    \int_p \! \int_k \! 
    {q_-^2 \over [\nu(q_+) q_+^2 + \nu(q_-) q_-^2]^2} \nonumber \\
  &&\qquad \qquad \times {1 \over \nu(q_+) q_+^2 +
    \nu({\bar q}_+) {\bar q}_+^2 + \nu({\bar q}_-) {\bar q}_-^2} \ ,
\label{2loop_prop}
\end{eqnarray}
while the result for the two--loop contribution due to vertex corrections
(Fig.~\ref{2loop_diagrams}g--j) is
\begin{eqnarray}
  &&\Gamma_{{\tilde h}h}^{(2,{\rm v})}(q,0) = - q^2 \lambda^4
    \int_p \! \int_k \! {q_- \over \nu(q_+) q_+^2 + \nu(q_-) q_-^2} 
\label{2loop_vertex} \\
  &&\times {{\tilde q}_+ \over
    [\nu({\tilde q}_+) {\tilde q}_+^2 + \nu({\tilde q}_-) {\tilde q}_-^2]
    [\nu(q_+) q_+^2 + \nu({\tilde q}_-) {\tilde q}_-^2 + \nu(k) k^2]} \ .
\nonumber
\end{eqnarray}
%

%%%%%%%%%% Mode Coupling Theory %%%%%%%%%%%%%%%%%%%%%%%%%%%%%%%%%%%%%%%%%%%%%%%

\section{Mode coupling theory and vertex corrections}
\label{mc_theory}

In this section we study the mode coupling approximation for the Burgers
equation. For readers not familiar with the dynamical functional approach
discussed in the previous section, we start by a derivation of the mode
coupling equation using a perturbation theory for the equation of motion.

%%%%%%%%%%%%%%%%%%%%%%%%%%%%%%%%%%%%%%%%%%%%%%%%%%%%%%%%%%%%%%%%%%%%%%%%%%%%%%%

\subsection{Perturbation series and mode coupling equations}
\label{mc_a}

In Fourier space with the equation of motion (\ref{kpz}) reads
\begin{eqnarray}
 &&h({\bf k},\omega) = 
 G_0({\bf k},\omega)\eta({\bf k},\omega) 
 + G_0({\bf k},\omega) {\widetilde{j}}({\bf k},\omega)\nonumber \\
 &&+ \frac{1}{2} G_0({\bf k},\omega)
                 \int_{{\bf q},\mu} V^{(0)}_{{\bf k}_+;{\bf k}_-}
                 h({\bf k}_+,\omega_+) h({\bf k}_-,\omega_-) \ . 
\label{kpzf}
\end{eqnarray}
where $G_0({\bf k},\omega) = 1 / (\nu {\bf k}^2 -i \omega)$ is the ``bare
propagator'', $V^{(0)}_{{\bf k}_1;{\bf k}_2} = - \lambda \, {\bf k}_1 \cdot
{\bf k}_2$ is the ``bare vertex'', and ${\bf k}_\pm \equiv
{\bf k}/2 \pm {\bf q}$, $\omega_\pm \equiv \omega/2 \pm \mu$. The noise $\eta$
is assumed to be Gaussian and uncorrelated, given by the weight function
Eq.~(\ref{weight_function}). A small external perturbation
${\widetilde{j}}({\bf k},\omega)$ has also been included in Eq.~(\ref{kpzf}),
and will be used to generate the response functions. Typically, the quantities
of interest are the noise--averaged two--point correlation function
\begin{equation}
  \langle h({\bf K})h({\bf K}') \rangle = C({\bf K})
  \delta ({\bf K}+{\bf K}') \ ,
\label{C}
\end{equation}
and the noise--averaged linear response function
\begin{equation}
  \bigg \langle \frac{\delta h({\bf K})}
  {\delta{\widetilde{j}}({\bf K}')} \bigg \rangle
  = G({\bf K}) \delta ({\bf K}-{\bf K}') \ ,
\label{G}
\end{equation}
where $({\bf k},\omega)$ is abbreviated by ${\bf K}$ and
$\delta ({\bf K}+{\bf K}') = (2\pi)^{d+1} \delta^d({\bf k}+{\bf k}')
\delta(\omega+\omega')$. These are special cases of the general Green's
function
\begin{eqnarray}
  &&G_{m,n}(-{\bf P}_1;\ldots;-{\bf P}_m|{\bf K}_1;\ldots;{\bf K}_n) 
  \nonumber \\
  &&\qquad = \bigg \langle \frac{\delta^m ( h({\bf K}_1)\cdots h({\bf K}_n))}
    {\delta{\widetilde{j}}({\bf P}_1) \cdots\delta{\widetilde{j}}({\bf P}_m)} 
    \bigg \rangle_c \ ,
\label{green} 
\end{eqnarray}
where the subscript ``c'' denotes the connected part. In this notation, the
two--point correlation function is $\langle h({\bf K})h({\bf K}') \rangle =
G_{0,2}(\cdot|{\bf K};{\bf K}')$ and the linear response function is $\langle
\delta h({\bf K})/\delta {\widetilde{j}}({\bf K}') \rangle =
G_{1,1}(-{\bf K}'|{\bf K})$. From Eq.~(\ref{green}), it is clear that
$G_{m,0} = 0$. The above definition of the Green's functions is identical to
the one used in the dynamic functional formalism (Sec.~\ref{rg_a}).

One approach to study the Green's functions $G_{m,n}$ is perturbation theory.
For $V^{(0)}=0$, Eq.~(\ref{kpzf}) is just the linear diffusion equation. For
$V^{(0)} \neq 0$, the solution of Eq.~(\ref{kpzf}) may be obtained iteratively
by a perturbation expansion in powers of $V^{(0)}$. For example, the lowest
order correction to the response function is
\begin{equation}
  G_1({\bf K}) 
  = G_0({\bf K}) + G_0({\bf K})\Sigma_1({\bf K}) G_0({\bf K}) \ ,
\label{G1} 
\end{equation}
where $C_0({\bf K}) = 2D |G_0({\bf K})|^2$ is the ``bare correlator'', and
\begin{equation}
   \Sigma_1({\bf K}) =  - \int_{{\bf Q}} 
   V^{(0)}_{{\bf k}_+;{\bf k}_-} G_0({\bf K}_-) C_0({\bf K}_+) 
   V^{(0)}_{{\bf k}_+;{\bf k}}
\label{sigma1} 
\end{equation}
is the one--loop renormalization of the ``self--energy''. Similarly, the lowest
order correction to the correlation function is
\begin{equation}
  C_1({\bf K})  =  2D_1({\bf K}) |G_0({\bf K})|^2 \ ,
\label{C1} 
\end{equation}
where
\begin{equation}
    D_1({\bf K}) = D + \frac{1}{4} \int_{{\bf Q}}  
    V^{(0)}_{{\bf k}_+;{\bf k}_-} C_0({\bf K}_-) C_0({\bf K}_+) 
    V^{(0)}_{{\bf k}_+;{\bf k}_-}
\label{D1} 
\end{equation}
is the one--loop renormalization of the ``noise spectrum''. Unfortunately, such
perturbation series diverge in the hydrodynamic limit ${\bf k},\omega \to 0$.
One way to proceed is to perform a renormalization group analysis. It turns
out, however, that there is no fixed point that can be obtained in a controlled
$\varepsilon$ expansion with $\varepsilon = 2-d$ below $d=2$
dimensions~\cite{ft94}. Hence, a non--perturbative method is required to treat
the KPZ problem. One approximation which has been frequently used is to replace
the bare propagator $G_0$ and bare correlator $C_0$ in Eqs.~(\ref{sigma1}) and
(\ref{D1}) by the renormalized functions $G$ and $C$ while keeping the vertex
$V^{(0)}$ unchanged. This is known as the mode--coupling approximation (or
Kraichnan's Direct Interaction Approximation), and it leads to the following
closed set of integral equations,
\begin{eqnarray}
       &\Sigma({\bf K}) &= - \int_{{\bf Q}} V^{(0)}_{{\bf k}_+;{\bf k}_-} 
                    G({\bf K}_-) C({\bf K}_+) V^{(0)}_{{\bf k}_+;{\bf k}} \ ,
\label{mc1}\\
       &D({\bf K}) &= D + \frac{1}{4} \int_{{\bf Q}} 
        V^{(0)}_{{\bf k}_+;{\bf k}_-}
        C({\bf K}_-) C({\bf K}_+) V^{(0)}_{{\bf k}_+;{\bf k}_-} \ ,
\label{mc2} 
\end{eqnarray}
where $\Sigma$ and $D$ are defined by $G$ and $C$ through
\begin{eqnarray}
  &G^{-1}({\bf K}) &= G_0^{-1}({\bf K}) - \Sigma({\bf K}), \label{sigma}\\
  &C({\bf K}) &=  2D({\bf K}) |G({\bf K})|^2 \ . \label{D} 
\end{eqnarray}
Of course, as this procedure neglects any vertex renormalizations, it
constitutes a partial sum of the perturbation series only, and as a--priori no
information is available about the size of the missing contributions, it
clearly constitutes an uncontrolled approximation. Nevertheless, the mode
coupling theory has been quite successfully applied in many areas of condensed
matter theory, as mentioned in the introduction. It was first applied to the
KPZ problem by van Beijeren, Kutner, and Spohn~\cite{dds} to get the scaling
exponents $\chi$ and $z$ in $d=1$. Recently, the mode--coupling equations were
solved numerically to obtain the entire function $C({\bf k},\omega)$ in $1+1$
dimensions \cite{hf91}, and striking agreement to the scaling function obtained
by direct numerical simulations~\cite{kmh92} were found (for details see
Sec.~\ref{mc_c} below). This result is very surprising and prompted us to study
the mode--coupling theory in more detail. In what follows, we will try to give
a systematic analysis of the mode--coupling approach using the
field--theoretic formulation of Langevin dynamics.

%%%%%%%%%%%%%%%%%%%%%%%%%%%%%%%%%%%%%%%%%%%%%%%%%%%%%%%%%%%%%%%%%%%%%%%%%%%%%%%

\subsection{Dynamic field theory, Ward identities, and vertex corrections}
\label{mc_b}

The starting point of our study is the response function, which can
be formally obtained by differentiating Eq.~(\ref{kpzf}) with respect to
the perturbation ${\widetilde{j}}({\bf K}')$. We obtain
\begin{eqnarray}
  &&G_{1,1}(-{\bf K}'|{\bf K}) = G_0({\bf K})\delta({\bf K}-{\bf K}')
         \nonumber \\
  &&+ \frac{1}{2} G_0({\bf K}) \int_{{\bf Q}} V^{(0)}_{{\bf k}_+;{\bf k}_-} 
                       G_{1,2}(-{\bf K}'|{\bf K}_+;{\bf K}_-) \ . 
\label{mc0}
\end{eqnarray}
Mode coupling theory amounts to expressing $G_{1,2}$ in terms of the lower 
order functions $G_{1,1}$ and $G_{0,2}$. 

In order to analyze the Green's functions systematically, we turn to a
functional integral method described in section \ref{rg_theory}. From the
generating functional, Eq.~(\ref{generating_functional}) the Green's functions,
Eq.~(\ref{green}), can be easily obtained as the functional derivatives of
$F[ {\widetilde{j}},j] = \log Z [ {\widetilde{j}},j ]$.

By taking derivatives of Eqs.~(\ref{def_fields}) and using
Eq.~(\ref{legendre}), it is straightforward to relate the Green's functions
$G_{m,n}$ to the vertex function $\Gamma_{m,n}$'s, e.g.,
\begin{equation}
  \Gamma_{1,1}({\bf P}|{\bf K}) = \Gamma_{1,1} (-{\bf P}) =
  G^{-1}({\bf P}) \delta({\bf K}+{\bf P}) \ .
\label{gamma11}
\end{equation}
The two--point correlation function can also be easily found. It has the form
of Eq.~(\ref{D}), with
\begin{equation}
  \Gamma_{2,0}({\bf P}_1,{\bf P}_2|\cdot) = \Gamma_{2,0} ({\bf P}_1) 
  = -2D({\bf P}_1) \delta({\bf P}_1+{\bf P}_2) \  .
\label{gamma20}
\end{equation}
\end{multicols}
\widetext
\noindent
All higher order Green's functions can be written as products of
$G({\bf K})$, $C({\bf K})$, and the higher order vertex functions. For example,
\begin{eqnarray}
  G_{2,1}({\bf P}_1;{\bf P}_2|{\bf K}) &= 
  &-G({\bf K})
  \, \Gamma_{1,2}({\bf K}|{\bf P}_1;{\bf P}_2)
  \, G(-{\bf P}_1)G(-{\bf P}_2) \ ,
  \label{G21}\\
  G_{1,2}({\bf P}|{\bf K}_1;{\bf K}_2) &=
  &-G({\bf K}_1)G({\bf K}_2) 
    \, \Gamma_{2,1}({\bf K}_1;{\bf K}_2|{\bf P}) 
    \, G(-{\bf P})\nonumber\\
  &     &-G({\bf K}_1)C({\bf K}_2) 
          \, \Gamma_{1,2}({\bf K}_1|{\bf K}_2;{\bf P}) 
          \, G(-{\bf P})\label{G12}\\
  &     &-C({\bf K}_1)G({\bf K}_2) 
         \, \Gamma_{1,2}({\bf K}_2|{\bf K}_1;{\bf P}) 
         \,  G(-{\bf P}) \ . \nonumber 
\end{eqnarray}
Using Eq.~(\ref{G12}) in Eq.~(\ref{mc0}), we obtain
\begin{eqnarray}
     &G_{1,1}(-{\bf K}'|{\bf K}) &= G({\bf K}) 
      \delta({\bf K}-{\bf K}')\nonumber \\
     &   &= G_0({\bf K}) \delta({\bf K}-{\bf K}') 
        - G_0({\bf K})\int_{{\bf Q}} V^{(0)}_{{\bf k}_+;{\bf k}_-} 
          G({\bf K}_-) 
         \nonumber \\
     &   & \times  \left[ C({\bf K}_+) 
         \Gamma_{1,2}({\bf K}_-|{\bf K}_+;-{\bf K}') 
        + \frac{1}{2} G({\bf K}_+) 
          \Gamma_{2,1}({\bf K}_-;{\bf K}_+|-{\bf K}') \right]
         G({\bf K}') \label{mc'} 
\end{eqnarray}
\begin{multicols}{2}
\narrowtext
\noindent
for the full response function. Note that it has the form $G({\bf K}) =
G_0({\bf K}) + G_0({\bf K})\Sigma({\bf K}) G({\bf K})$, where $\Sigma({\bf K})$
is the self energy defined in Eq.~(\ref{sigma}). If we write the vertex
functions as
\begin{eqnarray}
      &\Gamma_{1,2}({\bf K}|{\bf K}_1;{\bf K}_2) &= 
       \Gamma_a({\bf K}_1;{\bf K}_2) 
       \delta({\bf K} + {\bf K}_1 + {\bf K}_2) \ , \label{ga} \\
       &\Gamma_{2,1}({\bf P}_1;{\bf P}_2|{\bf K}) &= 
        \Gamma_b({\bf P}_2;{\bf K}) 
        \delta({\bf K} + {\bf P}_1 + {\bf P}_2) \label{gb} \ , 
\end{eqnarray}
then the self--energy  becomes
\begin{equation}
    \Sigma({\bf K}) = 
  - \int_{{\bf Q}} V^{(0)}_{{\bf k}_+;{\bf k}_-} 
    G({\bf K}_-) C({\bf K}_+) V({\bf K}_+; {\bf K}) \ , 
\label{mc}
\end{equation}
where
\begin{eqnarray}
  V({\bf K}_+;{\bf K})  = &&\Gamma_a({\bf K}_+;-{\bf K}) \nonumber \\
  &&+ \frac{G^{-1}(-{\bf K}_+)}{4D({\bf K}_+)}  \Gamma_b({\bf K}_+;-{\bf K}) 
\end{eqnarray}
denotes the ``renormalized vertex function''. It will be useful to write the
vertex function in a slightly different form,
\begin{eqnarray}
    V({\bf K}_+; &&{\bf K})  =  
    \Gamma_a({\bf K}_+;-{\bf K}) \nonumber \\
  + &&\frac{G^{-1}({\bf K}_+) + G^{-1}(-{\bf K}_+)}{4D({\bf K}_+)}  
                      \Gamma_b({\bf K}_+;-{\bf K}) \ .
\label{V}
\end{eqnarray}
The additional term does not change $\Sigma({\bf K})$ in Eq.~(\ref{mc}) because
its poles, from $G({\bf K}_-)$ and $G(-{\bf K}_+)$, are on the same side of the
complex frequency plane. Hence the frequency integral for this additional term
yields zero. Comparing Eq.~(\ref{mc}) with Eq.~(\ref{mc1}), we realize that the
mode--coupling equation becomes exact if $V({\bf K}_+;{\bf K}) =
V^{(0)}_{{\bf k}_+;{\bf k}}$. In section~\ref{mc_d}, we will show that this
equality in fact does not hold. Yet, by exploiting a number of identities
relating the different vertex functions, we shall show that the correction to
$V({\bf K}_+;{\bf K})$ is small in the limit ${\bf K}\to 0$. This is hopefully
the first step in understanding the puzzle of why the mode coupling theory
works so well, at least in the case of the noisy Burgers equation.

%%%%%%%%%%%%%%%%%%%%%%%%%%%%%%%%%%%%%%%%%%%%%%%%%%%%%%%%%%%%%%%%%%%%%%%%%%%%%%%

\subsection{Numerical solution of the mode coupling equations}
\label{mc_c}

In this section we present the numerical solution of the (1+1)--dimensional KPZ
equation. In view of the results from the preceding section it is convenient to
define a generalized kinetic coefficient $D(k,\omega)$ and a generalized
``surface tension'' $\nu (k,\omega)$ by
\begin{equation}
  G(k,\omega) = { 1 \over - i  \omega +  \nu(k,\omega)} \ ,
\label{mc_g}
\end{equation}
\begin{equation}
  C(k,\omega) = 
  {2 D(k,\omega) \over \omega^2 + [\nu(k,\omega)]^2} \ .
\label{mc_C}
\end{equation}
Then, the self-consistent equations for the correlation function 
$C(k,\omega)$ and the response function $G(k,\omega)$ in Fourier space are
given by
\begin{eqnarray}
  \nu(k,\omega) = &&\lambda^2 \int_{q,\mu} 
    k_+^2 k k_- C(k_+,\omega_+) G(k_-,\omega_-) \ ,
\label{mc_nu} \\ 
  D(k,\omega) = &&{\lambda^2 \over 4} \int_{q,\mu} 
    k_+^2 k_-^2 C(k_+,\omega_+) C(k_-,\omega_-) \ . 
\label{mc_D}
\end{eqnarray}
For the numerical solution of the mode--coupling equations it is much more
convenient to study the intermediate correlation and response functions, 
defined by full or half--sided Fourier transforms, respectively,
\begin{eqnarray}
  C(k,\omega) = &&\int_{-\infty}^{\infty} dt e^{i \omega t} C(k,t) \ , \\
  G(k,\omega) = &&\int_{-\infty}^{\infty} dt e^{i \omega t} \Theta (t) 
                                                            G(k,t) \ .
\end{eqnarray}
Inserting into Eqs.~(\ref{mc_nu})--(\ref{mc_D}) we get for the generalized
kinetic coefficients
\begin{eqnarray}
  \nu(k,t) = &&\lambda^2 \int_q 
    k_+^2 k k_- C(k_+,t) G(k_-,t) \ , \\ 
  D(k,t) = &&\left( {\lambda \over 2} \right)^2 \int_q 
    k_+^2 k_-^2 C(k_+,t) C(k_-,t) \ ,
\end{eqnarray}
where 
\begin{eqnarray}
  D(k,\omega) = &&\int_{-\infty}^{\infty} dt e^{i \omega t} D(k,t) \ , \\
  \nu(k,\omega) = &&\int_{-\infty}^{\infty} dt e^{i \omega t} \Theta (t) 
                                                              \nu(k,t) \ .
\end{eqnarray}
It is again important to realize that for the (1+1)--dimensional KPZ equation
there exists a fluctuation--dissipation theorem (FDT) which relates the
generalized kinetic coefficient $D(k,t)$ with the generalized surface tension
$\nu(k,t)$. As shown in Appendix A (see Eq.~(\ref{a12})), the following
identity holds
\begin{equation}
  G(k,t) = {\nu k^2 \over D} \Theta (t) C (k,t) \ .
\end{equation}
This allows one to rewrite $\nu(k,t)$ as
\begin{equation}
  \nu(k,t) =  k^2 {\lambda^2 D \over 2 \nu} 
              \int_q G(k_+,t) G(k_-,t) \, .
\end{equation}
Together with Eq.~(\ref{mc_g}), which can be written as
\begin{equation}
 {\partial \over \partial t} G(k,t) = 
 - \int_0^t d \tau \nu(k,\tau) G(k,t-\tau)\, ,
\end{equation}
one now has a set of self--consistent equations for the effective surface
tension and the response function.

\noindent{\em Scaling analysis of the mode coupling equations}.
We look for the solutions of the scaling form 
\begin{eqnarray}
  \nu(k,\omega) = &&{\bar \lambda} \nu k^z {\hat \nu} ({\hat \omega}) \ ,
\label{scaling_omega_nu} \\
  D(k,\omega) = &&{\bar \lambda} D k^{-\mu }{\hat \nu} ({\hat \omega}) \ ,
\label{scaling_omega_D}
\end{eqnarray}
where we have defined the scaling variable ${\hat \omega} = 
{\omega / {\bar \lambda} \nu k^z}$. The corresponding scaling forms for the
Fourier--transformed quantities read
\begin{eqnarray}
  \nu(k,t) = &&\left({\bar \lambda} \nu k^z\right)^2 {\hat \nu} ({\hat t}) \ ,
\label{scaling_time_nu} \\
  D(k,t) =  &&{\bar \lambda}^2 D \nu k^{z-\mu} {\hat \nu} ({\hat t}) \ ,
\label{scaling_time_D}
\end{eqnarray}
with the scaling variable ${\hat t} = {\bar \lambda} \nu k^z t$. For the
response function the scaling analysis leads to
\begin{eqnarray}
  G(k,\omega) = &&{1 \over {\bar \lambda} \nu k^z} {\hat G} ({\hat \omega}) 
  \ , \\
  G(k,t) = && {\hat G} ({\hat t}) \ .
\end{eqnarray}
Inserting the scaling forms,
Eqs.~(.\ref{scaling_omega_nu})--(\ref{scaling_time_D}), into the mode--coupling
equations implies for the dynamic exponent $z=3/2$, and leads to the following
self--consistency equations for the generalized kinetic coefficient and the
response function:
\begin{equation}
  {\hat \nu} ({\hat t}) 
  = {1 \over 2 \pi} \int_0^\infty dx {\hat G} (x_+ {\hat t})
                                     {\hat G} (x_- {\hat t}) \ ,
\label{mc_scal_nu}
\end{equation}
\begin{equation}
 {\partial \over \partial t} {\hat G} ({\hat t}) = 
 - \int_0^{\hat t} d {\hat \tau} {\hat \nu} ({\hat \tau}) 
   {\hat G} ({\hat t}-{\hat \tau}) \ ,
\label{mc_scal_G}
\end{equation}
where $x_\pm = 1/2 \pm x$, and the effective coupling constant is given by
\begin{equation}
  \bar\lambda^2 = \lambda^2 D/\nu^3 \ .
\end{equation}
Note that the amplitude ${\bar \lambda}$ is arbitrary. We have chosen it to be
equal to the effective coupling constant in order to simplify the scaled
mode--coupling equations.

\begin{figure}
\narrowtext
\epsfysize=2.8truein 
\epsfysize=\columnwidth{\rotate[r]{\epsfbox{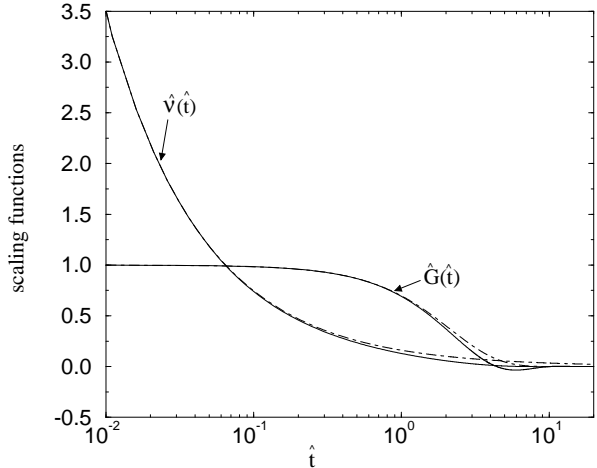}}}
\vspace{15pt}
\caption{Scaling functions for the generalized surface tension 
  ${\hat \nu} ({\hat t})$ and response function ${\hat G} ({\hat t})$ vs. the
  scaling variable ${\hat t} = \lambda (D / \nu)^{1/2} k^{3/2} t$. The
  point--dashed line represents the Gaussian approximation,
  Eq.~(\protect\ref{gauss_app}), for the response function, which is obtained
  from the analysis of the mode coupling equations at small times.}
\label{scaling_functions}
\end{figure}

It can be shown analytically from Eq.~(\ref{mc_scal_nu}) that the scaling
function for the generalized surface tension ${\hat \nu} ({\hat t})$ shows a
power--law behavior ${\hat \nu} ({\hat t}) = {\bar \nu} {\hat t}^{-2/3}$ with
${\bar \nu} \approx 0.1608$ for small times ${\hat t} \leq 10^{-1}$. Since the
response function is almost constant for small $\hat{t}$, one finds from
Eq.~(\ref{mc_scal_G}) that
\begin{equation}
  {\hat G} ({\hat t}) = 
  \exp \left\{ - C_{\rm gauss}  {\hat t}^{4/3}  \right\} \, \, 
  {\rm for} \, \, {\hat t} \leq 10^{-1} \ .
\label{gauss_app}
\end{equation}
with $C_{\rm gauss} = 9 {\bar \nu} / 4 \approx 0.3619$. In
Fig.~\ref{scaling_functions} the numerical solutions for the scaling functions 
${\hat \nu} ({\hat t})$ and ${\hat G} ({\hat t})$ are depicted, as well as the
results from the Gaussian approximation (\ref{gauss_app}).

\begin{figure}
\narrowtext
\epsfysize=\columnwidth{\rotate[r]{\epsfbox{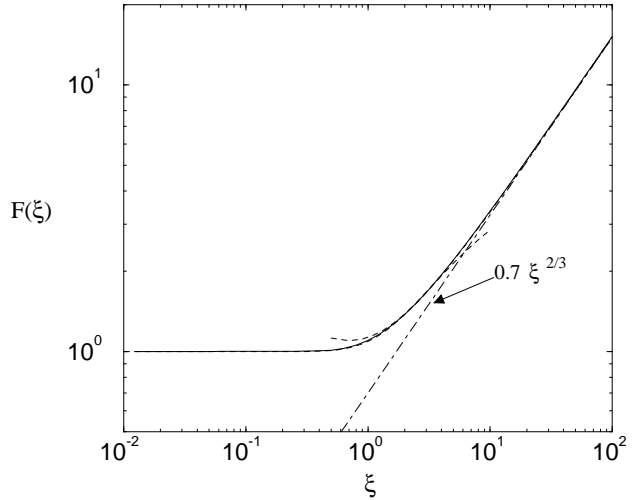}}}
\vspace{15pt}
\caption{Scaling function $F(\xi)$ for the truncated correlation function in
  real space versus the scaling variable $\xi= \lambda A^{1/2} t/ x^{3/2}$. The
  empirical forms, Eq.~(\protect\ref{empirical_form}), are shown as the dashed
  curves. The dimensionless coupling constant can be read off from the
  crossover point of $F(\xi)$.}
\label{scalrealspace}
\end{figure}

\noindent {\em Truncated correlation function in real space}.
Another quantity, which is easily accessible by numerical simulations, is the
{\it truncated} correlation function in real space,
\begin{eqnarray}
  C(x,t) 
  && = \int_{-\infty}^{\infty} {dk \over 2\pi}
      2 \left[ 1 - e^{ikx} \right] 
      C(k,t) \nonumber \\
  && = {D \over \nu} x F 
      \left( {\lambda \sqrt{D/\nu}\>t \over x^{3/2}} \right) 
      \ ,
\end{eqnarray}
where $F(0) = 1$ since $G(0) = 1$. Conforming to the definition in
Eq.~(\ref{invariance_scaling}) one gets $A = D/\nu$.  The universal scaling
function $F(\xi)$ and is shown in Fig.~\ref{scalrealspace}.
The dimensionless argument of $F$ has the form demanded by
Eq.~(\ref{invariance_scaling}) with $z=3/2$. The dimensionless coupling
constant can be read off from the crossover point of $F(\xi)$ (see
Fig.~\ref{scalrealspace}); we obtain $g^{*} = 0.87$. This result can be checked
more precisely in simulations by directly looking at the scaling amplitudes. 
Our work thus predicts that if
\begin{equation}
  C(x,t=0) = A x \ ,
\label{C_x}
\end{equation}
then
\begin{equation}
  C(x=0,t) = 0.70 (\lambda A^2 t) ^{2/3} \  . 
\label{C_t}
\end{equation}
The numerical error is less than $\pm 1\%$. The amplitude $0.70 \pm 0.1$,
extracted from the mode coupling equations, agrees rather well with results
from numerical simulations, which find an amplitude of
$0.712 \pm 0.003$~\cite{kmh92} and $0.725 \pm 0.005$~\cite{t92}, respectively. 
In Ref.~\cite{t92} a empirical form for $F(\xi)$ has been given, which fits the
data from the numerical simulation quite well. We find that the mode--coupling
result is also quite well approximated by the same empirical forms (dashed
curves in Fig.~\ref{scalrealspace})
\begin{equation}
  F(\xi) = \cases{1\!+\!4.22 \exp \left\{ -3.82 \xi^{3/2} \right\} 
                  & {\rm for} $\xi \leq \xi_0$, \cr  
                  0.7 \xi^{2/3}\!+\!0.43 \xi^{-2/3} & {\rm for}
                  $\xi \geq \xi_0$,}
\label{empirical_form}
\end{equation}
where $\xi_0 \approx 2.5$ (see the dashed curves in Fig.~\ref{scalrealspace}),
but with somewhat different numerical values for the coefficients. Note that
the dashed curves are almost indistinguishable from the solid line; in order to
make the dashed curves visible we have plotted the asymptotic forms in
Eq.~\ref{empirical_form} for values smaller and larger than $\xi_0$,
respectively. In summary, the mode coupling result for the scaling function
$F(\xi)$ agrees with the results from numerical simulation~\cite{kmh92,t92}
within a few percent.

\noindent{\it Finite size effects}. 
Note that the above results are valid for very large systems in the {\em steady
state}. Transient behaviors such as the growth of interfacial width starting
from flat initial conditions may well be more complicated~\cite{family90}. They
may also be computed using the mode-coupling theory (with Fourier--Laplace
transform to incorporate the initial conditions; however, the procedure becomes
more cumbersome.

Nevertheless, we can say something about the behavior of systems of finite size
$L$ already on the basis of our results for the steady state. In principle, the
correlation function and response function are now explicitly $L$--dependent.
They may be described in terms of $D_L(k,\omega)\sim\nu_L(k,\omega)\sim
L^{1/2}f(\hat\omega, kL)$, where $f$ is now the solution of a two variable
integral equation with the initial condition $D_{L=a}(k,\omega)=D$ (the bare
value) and similarly for $\nu$. In this way, one would obtain the explicit {\em
functional} renormalization of various quantities as we look at larger length
scales $L$. The flow behavior of $D$ and $\nu$ described by usual recursion
relations are recovered from the $L$ dependence of $D_L(k=0,\omega=0)$ and
$\nu_L(k=0,\omega=0)$. The asymptotic form $D_L\sim\nu_L\sim L^{1/2}$ is of
course the expected one given the exponents $\chi$ and $z$\cite{nu}. Here we
want to emphasize that the self-consistent equations provide a connection
between the microscopic and macroscopic (renormalized) theory.

If the flow of these functions is already well advanced, i.e., for times much
larger than the initial time $t=0$, where the interface was absolutely flat,
our results for the steady state can also be used to get approximate results
for the ``transient behavior''of a finite size system.  Note that with
``transient behavior'' we are not referring to the transients starting out from
an absolutely flat interface, but to transient behavior after some initial
relaxation.

\begin{figure}
\narrowtext
\epsfysize=\columnwidth{\rotate[r]{\epsfbox{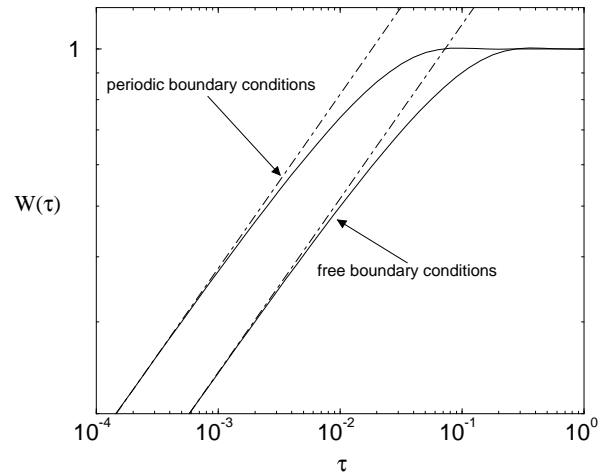}}}
\vspace{15pt}\caption{Scaling function $W(\tau)$ for the interface width in a
  system of finite size $L$ for periodic boundary conditions and free boundary
  conditions, respectively. The dashed lines are approximation for small scaled
  times $\tau = \lambda C_L t L^{-3/2}$, $W(\tau) \approx 3.8 \tau^{1/3}$ for
  periodic boundary conditions and $W(\tau) \approx 2.4 \tau^{1/3}$ for free
  boundary conditions.}
\label{scaling_width}
\end{figure}

The interface width in a system of finite size $L$ is defined by
\begin{equation}
  w^2_L (t) = \left< \left[ h(x,t) - h(x,0) \right]^2 \right> \bigg \vert_L \ .
\label{interface_width_def}
\end{equation}
Upon assuming that the spectrum of the height function is only slightly
modified by finite size effects (and/or after some initial transient), the
interface width can be approximated in terms of the correlation function in the
steady state
\begin{eqnarray}
   w^2_L (t) = &&2 \int_{-\infty}^{+\infty} {d \omega \over 2 \pi}
               \left( 1 - e^{i \omega t }\right)
               2 \int_{2 \pi /L}^{+\infty} {d k \over 2 \pi}
               C(k,\omega) \nonumber \\
             = &&{ 4 D \over \nu} 
               \int_{2 \pi /L}^{+\infty} {d k \over 2 \pi}
               \, {1 \over k^2} \, 
               \left[ 1 - G (k,t) \right]
\label{interface_width_app}
\end{eqnarray}
for periodic boundary conditions. Inserting the scaling laws
Eqs.~(\ref{scaling_omega_nu})--(\ref{scaling_time_D}) for the correlation
functions one finds
\begin{equation}
   w^2_L (t) =  {D \over \nu} L f ({\tilde \tau}) \ ,
\label{interface_width_scaling}
\end{equation}
where ${\tilde \tau } = \lambda \sqrt{{D \over \nu}} L^{-3/2} t$ and
\begin{equation}
   f ({\tilde \tau}) = 
   {2 \over \pi} \int_{2 \pi}^\infty d x x^{-2} 
   \left[ 1 - G ({\tilde \tau} x^{3/2} ) \right] \ .
\label{scaling_fun_interface_width}
\end{equation}
Asymptotically one gets
\begin{equation}
   w^2_L (t) = 
   \cases{ C_L^2 L       & for $t \to \infty$ \ , \cr
           C_t^2 t^{2/3} & for $t \to 0$ \ .} 
\label{asymptotics_interface_width}
\end{equation}
Just as in the discussion of the steady state correlation function one can
define an universal amplitude ratio by
\begin{equation}
   R = {C_t \over ( \lambda C_L^4 )^{1/3} } \ ,
\label{amplitude_L}
\end{equation}
and rewrite the scaling law in terms of this ratio
\begin{equation}
   w_L(t) = C_L \sqrt{L} W ( \tau ) \ ,
\label{interface_width_final_form}
\end{equation}
where $\tau = \lambda C_L t L^{-3/2}$, $C_L^2 = f (\infty) {D / \nu}$ with
$f(\infty) \approx 0.101$. The scaling function $W(\tau)$, shown in
Fig.~\ref{scaling_width}, has the following asymptotic behavior
\begin{equation}
   W(\tau) =
   \cases{ 1               & for $\tau \to \infty$ \ , \cr
           R \tau^{1/3}    & for $\tau \to 0$ \ .     }
\label{asymptotic_scaling_function}
\end{equation}
The ratio $R$ is found to be $R = 3.8$, which is in reasonable agreement with
$R_{\rm exp} = 3.45 \pm 0.05$ found in numerical simulations~\cite{af92}.
If one uses free instead of periodic boundary conditions, one has to replace
the lower bound of the integral in Eqs.~(\ref{interface_width_app}) and
(\ref{scaling_fun_interface_width}) by $\pi$. The resulting scaling function
for the interface width is also shown in Fig.~\ref{scaling_width}, and we find,
as already noted in Ref.~\cite{af92}, that the asymptotic behavior at small
times is given by $W ( \tau ) = R^f \tau^{2/3}$ with
$R^f = R/2^{2/3} \approx 2.4$.

%%%%%%%%%%%%%%%%%%%%%%%%%%%%%%%%%%%%%%%%%%%%%%%%%%%%%%%%%%%%%%%%%%%%%%%%%%%%%%%

\subsection{Vertex corrections}
\label{mc_d}

As we have seen in the preceding section mode coupling theory is equivalent to
a self--consistent formulation of the perturbation series, where all propagator
renormalizations are taken into account, but vertex corrections have been
neglected. Nevertheless, there is quite an excellent agreement of the mode
coupling results with numerical simulations~\cite{kmh92}. It seems that there
is some hidden small parameter, which remains to be identified. In this section
we address this problem and analyze the magnitude of the vertex corrections.

It is known that the KPZ equation is invariant under a Galilean transformation
of the form
\begin{eqnarray}
   h'({\bf x},t) &&= h({\bf x} + \lambda {\bf v} t, t) 
                     + {\bf v}  \cdot {\bf x} \ , \\
   {\widetilde{h}}'({\bf x},t) &&= 
             {\widetilde{h}}({\bf x} + \lambda {\bf v} t, t) \ ,
\end{eqnarray}
corresponding to an infinitesimal tilt ${\bf v}$ of the surface. This
invariance leads to Ward identities, connecting the two-- and three--point
vertex functions~\cite{ft94}, which imply that the nonlinearity $\lambda$ is
not renormalized, and that there is an exponent identity $\chi + z = 2$.

Since the Ward identities relate the three--point with the two--point vertex
functions one may hope that they also give some information on the magnitude of
the vertex functions. Recently, it has been shown by Lebedev and
L'vov~\cite{lebedev94} that the KPZ equation is invariant under the generalized
Galilean transformation
\begin{eqnarray}
   h'({\bf x}',t) 
   &&= h({\bf x},t) + {\partial \mbox{\boldmath $\zeta$} \over
                       \partial t} \cdot {\bf x}'  \, , \\
   {\widetilde{h}}'({\bf x}',t) 
   &&= {\widetilde{h}}({\bf x},t) \, ,
\end{eqnarray}
with ${\bf x}' = {\bf x} - \lambda \mbox{\boldmath $\zeta$}$, and where
$\mbox{\boldmath $\zeta$}$ is an arbitrary function of time but not of
coordinates ${\bf x}$. Since the generating functional for the vertex functions
$\Gamma [{\widetilde{h}}, h]$ is invariant with respect to the above
transformation, one finds the following Ward identity
\begin{eqnarray}
  \int_{\bf k} \int dt \Biggl[
  &&\lambda {\bf k} \cdot \mbox{\boldmath $\zeta$}
  \biggl\{ {\delta \Gamma \over \delta h({\bf k}, t)} h({\bf k}, t)
  + {\delta \Gamma \over \delta {\widetilde{h}}({\bf k}, t)} 
  {\widetilde{h}}({\bf k}, t) \biggr\} \nonumber \\
  + &&{\delta \Gamma \over \delta h({\bf k}, t)} 
  {\partial \mbox{\boldmath $\zeta$} \over \partial t} \cdot 
  {\partial \over \partial {\bf k}} \delta^{(d)} ({\bf k}) \Biggr] = 0 \ .
\label{ward_identity_source}
\end{eqnarray}
Taking functional derivatives of the above equation with respect to
${\widetilde{h}} (-{\bf q}_1,-\mu_1)$ and $h(-{\bf k}_1,-\omega_1)$, then
taking the limit $h,{\widetilde{h}} \to 0$, and recalling the definition of the
vertex functions, we obtain the following Ward identity
\begin{eqnarray}
  i \omega \lim_{{\bf k}\rightarrow 0} && {\partial \over \partial {\bf k}}
  \Gamma_{1,2} ( {\bf q}_1, \mu_1 \mid  {\bf k}_1, \omega_1; {\bf k}, \omega)
  \nonumber \\ = \lambda 
  \biggl\{ &&{\bf q}_1 \Gamma_{1,1} ({\bf q}_1, \mu_1 + 
         \omega \mid {\bf k}_1, \omega_1) \nonumber \\ 
         + &&{\bf k}_1 \Gamma_{1,1} ({\bf q}_1, \mu_1 \mid {\bf k}_1, 
         \omega_1 + \omega )
  \biggr\} \ .
\label{ward1}
\end{eqnarray}
Similarly, by taking derivatives of Eq.~(\ref{ward_identity_source}) with
respect to ${\widetilde{h}}(-{\bf q}_1,-\mu_1)$ and
${\widetilde{h}}(-{\bf q}_2,-\mu_2)$, we get
\begin{eqnarray}
  i \omega \lim_{{\bf k}\rightarrow 0} && {\partial \over \partial {\bf k}}
  \Gamma_{2,1} ( {\bf q}_1, \mu_1 ; {\bf q}_2, \mu_2 \mid {\bf k}, \omega)
  \nonumber \\ = \lambda 
  \biggl\{ 
  &&{\bf q}_1 \Gamma_{2,0} ({\bf q}_1, \mu_1 + \omega; 
  {\bf q}_2, \mu_2 \mid \cdot) \nonumber \\ + 
  &&{\bf q}_2 \Gamma_{2,0} ({\bf q}_1, \mu_1; 
  {\bf q}_2, \mu_2 + \omega \mid \cdot) 
  \biggr\} \, .
\label{ward2}
\end{eqnarray}
The general Ward identity reads
\begin{eqnarray}
  &&i \omega \bbox{\nabla}_{{\bf k}} 
  \Gamma_{m,n+1}( \{ {\bf Q}_i \} | \{ {\bf K}_i \}; 
                    {\bf k}, \omega ) |_{{\bf k} = 0} \nonumber \\
  = &&\lambda  \sum_{j=1}^m {\bf q}_j 
  \Gamma_{m,n} (  \{ {\bf Q}_i \} + \omega {\bf e}_j | \{ {\bf K}_i \} )
  \nonumber \\
  + &&\lambda \sum_{j=1}^n {\bf k}_j
  \Gamma_{m,n} ( \{ {\bf Q}_i \ | \{ {\bf K}_i \}  + \omega {\bf e}_j ) \ , 
\label{wardid}
\end{eqnarray}
where we have defined \hbox{$ \{ {\bf Q}_i \} = {\bf q}_1,\mu_1;\ldots;
{\bf q}_m,\mu_m$}, $ \{ {\bf K}_i \} = {\bf k}_1,\omega_1;\ldots;{\bf k}_{n},
\omega_{n}$, and $ \{ {\bf Q}_i \} + \omega {\bf e}_j = {\bf q}_1,\mu_1;\ldots;
{\bf q}_j,\mu_j + \omega ;\ldots;{\bf q}_m,\mu_m$. Inserting the above Ward
identities, Eq.~(\ref{ward1}) and Eq.~(\ref{ward2}), into the expression for
the vertex correction, we find
\begin{eqnarray}
   &&\bbox{\nabla}_{{\bf k}} V( {\bf K}_+; {\bf K})\mid_{{\bf k} = 0} =
   {\lambda \over i \omega} {\bf q} 
   \Biggl\{
    - i \omega \! + \! \nu^* ({\bf q},\mu_-) \! - \! 
                   \nu^* ({\bf q},\mu_+) \nonumber \\
    && - {\nu ({\bf q},\mu_+) \!
      + \! \nu^* ({\bf q},\mu_+)
       \over 2 D ({\bf q},\mu_+) }
    \left[ D ({\bf q},\mu_-) \! - \! 
           D ({\bf q},\mu_+)
    \right]  
   \Biggr\} \ ,
\label{vertex_correction}
\end{eqnarray}
where $\mu_\pm = \mu \pm \case \omega / 2$. The first term in the latter
equation corresponds to the bare vertex, which is real. The corrections to this
bare vertex result from the imaginary part of the frequency dependent surface
tension $\nu ({\bf k},\omega)$ in the second and third term of
Eq.~(\ref{vertex_correction}). In addition, the renormalized vertex contains an
imaginary part resulting from the real part of the generalized surface tension 
and the noise amplitude $D({\bf k}, \omega)$.
\begin{figure}
\narrowtext
\epsfysize=\columnwidth{\rotate[r]{\epsfbox{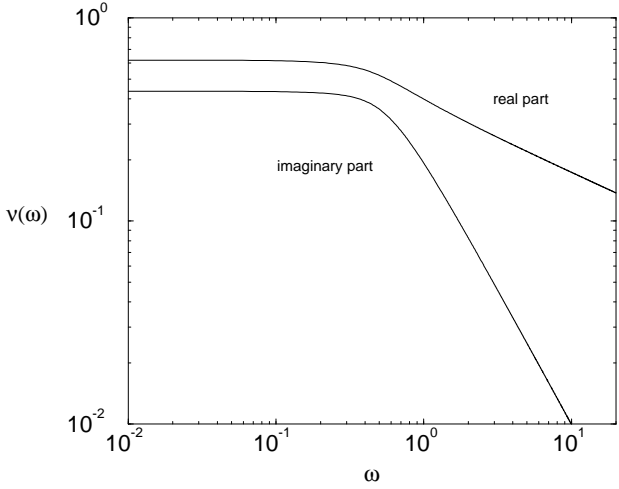}}}
\vspace{15pt}\caption{Scaling function for the real part,
  $Re \left[ {\hat \nu}({\hat \omega}) \right]$, and imaginary part divided by 
  the scaled frequency, $Im \left[ {\hat \nu}({\hat \omega}) \right] / 
  {\hat \omega}$, of the generalized surface tension.}
\label{surface_freq}
\end{figure}
Let us first discuss the vertex corrections resulting from the imaginary part
of the generalized surface tension. Using the mode coupling results from
Section \ref{mc_c}, one can calculate the real and imaginary part of $\nu
({\bf k}, \omega)$, as shown in Fig.~\ref{surface_freq}.
\begin{figure}
\narrowtext
\epsfysize=\columnwidth{\rotate[r]{\epsfbox{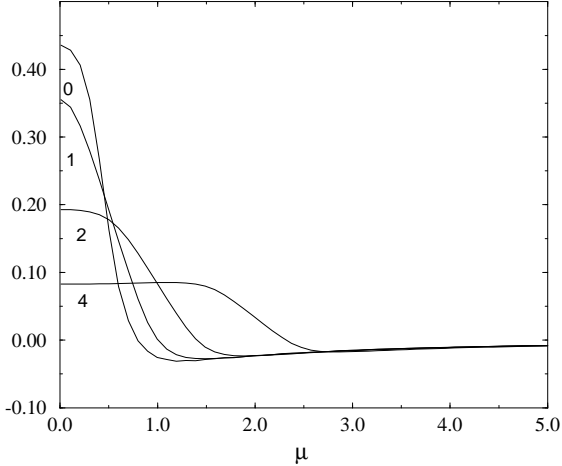}}}
\vspace{15pt}\caption{Real part of the vertex correction versus $\hat \mu$, 
  ${\rm Im} \left[ {\hat \nu} (\hat \mu - \case {\hat \omega} / 2) -
  {\hat \nu}(\hat \mu + \case {\hat \omega} / 2) \right] / {\hat \omega}$. The
  value of the scaling variable $\hat \omega$ is indicated in the graph.}
\label{vertex_real}
\end{figure}
Therefrom one deduces the real part of the vertex corrections
\begin{eqnarray}
   &&Re \left[ \bbox{\nabla}_{{\bf k}} V( {\bf K}_+; {\bf K})\mid_{{\bf k} = 0}
   \right]
   + {\lambda} {\bf q} = \\ \nonumber &&- {\lambda} {\bf q} 
   Im \left[ {\hat \nu} (\hat \mu - \case {\hat \omega} / 2) - 
             {\hat \nu} (\hat \mu + \case {\hat \omega} / 2) \right] / 
      {\hat \omega} \ ,
\end{eqnarray}
where $\hat \mu = \mu/{\bar \lambda} \nu q^{3/2}$ and
${\hat \omega} = \omega /{\bar \lambda} \nu q^{3/2}$. 
As can be inferred from Fig.~\ref{vertex_real} the vertex corrections may be as
large as $40\%$ at certain values of the external frequencies. If we take the 
integral over all frequencies as a measure of the vertex correction, however,
we find that it is only of the order of a few percent or even less.

%%%%%%%%%%%%%%%%%%%%%%%%%%%%%%%%%%%%%%%%%%%%%%%%%%%%%%%%%%%%%%%%%%%%%%%%%%%%%%%

\subsection{Vertex corrections from the two--loop contributions}
\label{mc_e}

\begin{figure}
\narrowtext
\epsfysize=\columnwidth{\rotate[r]{\epsfbox{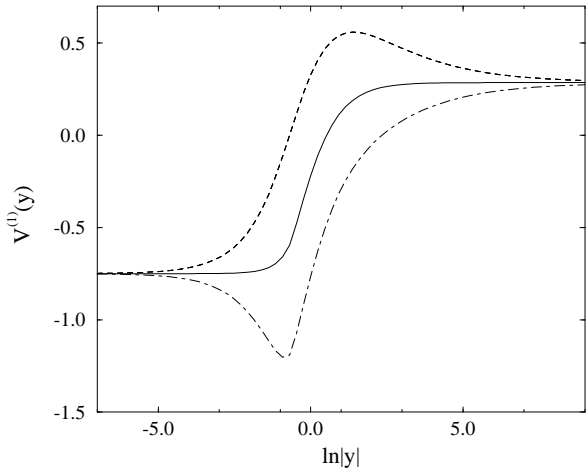}}}
\vspace{15pt}
\caption{Vertex correction $V^{(1)} (y)$ from two-loop diagrams as a function 
  of $\ln|y|$: dashed curve: $y < 0$, dot-dashed curve: $y>0$. The solid line
  represents the average ${\bar V}^{(1)} (y) = \left[ V^{(1)}(y) + V^{(1)}(-y)
  \right] / 2$.}
\label{vertex_corrections}
\end{figure}

In this subsection we study the vertex corrections resulting from two--loop
diagrams in Lorentzian approximation. With the ansatz $\nu(q) = \nu(q,\omega=0)
= \nu_{\rm lor} {(\lambda / \sqrt{2 \pi})} q^{z-2}$ the mode--coupling
equations in Lorentzian approximation read (note that $z=3/2$)
\begin{equation}
\nu_{\rm lor}^2 = {1 \over 2} \! \int_{-\infty}^{+\infty} \! dy  \!
                  {1 \over y_+^{3/2} \! + \! y_-^{3/2}}
\end{equation}
where $y_\pm = {\case 1/2} \pm y$. This gives $\nu_{\rm lor}^2 \approx 1.955$.
Next we take into account vertex corrections from the two--loop diagrams. We
have seen in section~\ref{rg_c} that the two--loop contributions to $\nu(q,0)$
can be split into a propagator renormalization and a vertex correction. The
former is already taken into account in the self--consistency scheme of mode
coupling theory. Hence, in Lorentzian approximation one may extend the
mode--coupling approach in the following way
\begin{equation}
\nu(q) = {\lambda^2 \over 2} \int_p {1 + V^{(1)} (p/q) \over
                                     \nu(q_+) q_+^2 + \nu(q_-) q_-^2} \ ;
\label{mc_vertex_corrected}
\end{equation}
with 
\begin{equation}
V^{(1)} (y) = {2 y - 1 \over \nu_{\rm lor}^2} 
              \int \! \! dx {{\case 1/2} + y + x \over
              \left[{\tilde y}_+^{3 \over 2}\!+\! {\tilde y}_-^{3 \over 2} 
              \right] 
              \left[{\tilde y}_-^{3 \over 2}\!+
                    \!y_+^{3 \over 2}\!+\!|x|^{3 \over 2} 
              \right]} \ ,
\end{equation}
where ${\tilde y}_\pm = |{\case 1/2} \pm y \pm x|$ and $y_\pm = |{\case 1/2}
\pm y|$.  As can be inferred from Fig.~\ref{vertex_corrections}, the vertex
corrections $V^{(1)}$ are not at all small as compared to the bare term $1$.
They vary from about $-100\%$ to $+100\%$ as a function of the ratio of the
external momenta $q$ and $p$.

However, by inserting this vertex correction in the extended mode coupling
equation, Eq.~(\ref{mc_vertex_corrected}), one finds
$\nu_{\rm lor}^2 \approx 2.044$, which is merely less than $5\%$ larger than
the value obtained with the bare vertex. In addition, the correction tends to
increase the amplitude ratio calculated in section \ref{mc_c} towards the value
obtained from numerical simulations. This result clearly explains why mode
coupling theory has been so successful in calculating scaling functions for the
noisy Bugers equation. 
It still remains a puzzle to us, however, that the vertex correction itself
constitutes such a large correction to the bare vertex of the order of
$\pm 100 \%$, while its effect on the correlation function is much less. Future
investigations may concentrate on the extension of the two--loop vertex 
correction to higher orders, e.g. by including all ladder diagrams to the 
three--point vertex function, which may give us further insight in the validity
and accuracy of the fairly simple mode--coupling approach.

%%%%%%%%%% Summary %%%%%%%%%%%%%%%%%%%%%%%%%%%%%%%%%%%%%%%%%%%%%%%%%%%%%%%%%%%%

\section{Summary and Conclusions}
\label{summary}

In this paper we have investigated the noisy Burgers equation using dynamical
renormalization group and mode--coupling techniques. The renormalization group
results show that there appears no singular two--loop contribution to the
height--height correlation function. Upon including the two--loop vertex
corrections into the mode--coupling approach we were able to show that their
effect on the result for the correlation function is approximately $5\%$,
whereas the vertex corrections themselves are quite large. We suppose that this
overestimates the actual vertex correction. In order to go beyond the two--loop
vertex corrections one should possibly use an additional suitable resummation
of the vertex correction, e.g., write down a ``Bethe--Salpeter'' type of
equation for the vertices representing all ladder diagrams contributing to the
three--point vertex functions. We leave this problem, and conceivable further
extensions, for future investigations.

As a complementary estimate for the vertex corrections we have used Ward
identities to derive a relation between certain derivatives of the three--point
vertex function and the renormalized noise amplitude and surface tension.
Again, one finds that the effect of the vertex corrections may be of the order
of a few percent. 

Based on the above estimates for the vertex corrections, we suppose that mode
coupling theory yields very accurate results for the scaling functions, at
least for the (1+1)--dimensional KPZ (noisy Burgers) equation. This conclusion
is supported by the close agreement of the mode coupling results with those 
from numerical simulations of finite size systems. It remains unclear, however,
whether the mode--coupling approach works as well in the general
$(d+1)$--dimensional case, where the two--loop perturbation theory corrections
do contain singular contributions.

\acknowledgements

E.F. and U.C.T gratefully acknowledge support from the Deutsche
Forschungsgemeinschaft (DFG) under Contract Nos. Fr.~850/2, SFB 266 and
Ta.~177/1, respectively. T.H. is supported by an A. P. Sloan Research
Fellowship and an ONR Young Investigator Award.

%%%%%%%%%% Appendix %%%%%%%%%%%%%%%%%%%%%%%%%%%%%%%%%%%%%%%%%%%%%%%%%%%%%%%%%%%

\appendix

\section{Generalized fluctuation dissipation relation}
\label{app_a}

In this appendix we collect several fluctuation dissipation theorems (FDT's),
which are of importance for the dynamics of systems described by nonlinear
Langevin equations. 

Let ${\cal T}$ be the time reversal operation: $t \rightarrow - t$.  Then
detailed balance (time inversion symmetry) implies~\cite{j79} that
\begin{equation}
  {\cal T} \exp \left[ {\cal J}_{t_1}^{t_2} - F_{t_1} \right] =
  \exp \left[ {\cal J}_{-t_2}^{-t_1} - F_{-t_2} \right] \ ,
\label{timereversal}
\end{equation}
where $F_t = F[h(t)]$ is the stationary probability distribution function and
\begin{eqnarray}
   {\cal J}_{t_1}^{t_2} [h,{\tilde h}] = 
   &&\int dx \int_{t_1}^{t_2} dt \Biggl[ {\tilde h} (x,t) D  
     {\tilde h} (x,t) \nonumber \\
   &&- {\tilde h} (x,t) \left( {\partial h (x,t) \over \partial t} 
     - V(h) \right) \Biggr] \ ,
\label{a2}
\end{eqnarray}
with
\begin{equation}
  V(h) = - D {\delta F \over \delta h} + 
           {\lambda \over 2} (\nabla h)^2 \ .
\label{a3}
\end{equation}
Without loss of generality, one can assume that the field $h(x,t)$ is even or
odd under time reversal, that is
\begin{equation}
  {\cal T} h = \varepsilon h, \quad \varepsilon = \pm 1 \ .
\label{a4}
\end{equation}
Here $h$ is odd under time reversal.  The stationary distribution $P_{\rm st}
[h] = e^{- F[h]}$ is characterized by the ``free energy''
\begin{equation}
  F[h] = {\nu \over 2 D} \int dx (\nabla h)^2 \ .
\label{a5}
\end{equation}
The time reversal symmetry implies that
\begin{equation}
  {\cal T} {\widetilde{h}} (t) =
  - \varepsilon \left( {\widetilde{h}} (-t) - 
   {\delta F [h(-t)] \over \delta h(-t)} \right) \ .
\label{a6}
\end{equation}
Now one uses the causality property of the response functions, $\langle h(t_1)
... h(t_k) {\widetilde{h}} ({\tilde t}_1) ... {\widetilde{h}} ({\tilde t}_k)
\rangle= 0$, if one ${\tilde t}_j \rangle \, {\rm all} \, t_i$. Then for
example
\begin{equation}
  \langle h(t) {\widetilde{h}} (0)\rangle  = 0 \, \, {\rm for} \, \, t < 0 \ .
\label{a7}
\end{equation}
With the time reversal operation and Eq.~(\ref{a6}) it follows then from
Eq.~(\ref{a7}) for $t<0$ that 
\begin{equation}
  \bigg \langle h(-t) \left( {\widetilde{h}} (0) - 
  {\delta F[h(0)] \over \delta h (0)} \right) \bigg \rangle  = 0 \ .
\label{a8}
\end{equation}
Upon redefining $t = - t$, one obtains for $t>0$
\begin{equation}
  \langle h(t) {\widetilde{h}} (0) \rangle  =  \Theta(t) 
  \bigg \langle h(t) {\delta F[h(0)] \over \delta h (0)} \bigg \rangle \ .
\label{a9}
\end{equation}
The same arguments can be repeated for $\langle h(t_1) ...
h(t_k){\widetilde{h}} ({\tilde t}_1) \rangle $ with 
${\tilde t}_1 > \, {\rm all} \, \, t_j$. The result is
\begin{eqnarray}
  &&\langle h(t_1)... h(t_k){\widetilde{h}} ({\tilde t}_1) \rangle 
  \nonumber \\ = 
  &&\Theta({\tilde t}_1,\{ t_j \} ) 
  \bigg \langle h(t_1) ... h(t_k) 
  {\delta F[h({\tilde t}_1)] \over \delta h ({\tilde t}_1)} \bigg \rangle \ .
\label{a11}
\end{eqnarray}
where $\Theta({\tilde t}_1,\{ t_j \} )$ is an obvious generalization of the
$\Theta$ function. Note that these generalized FDT's are for the cumulants and
not for the vertex functions. In particular we get
\begin{equation}
  G_{11} (k,t) = \Theta (t) { \nu k^2 \over D} G_{02} (k,t) \, .
\label{a12}
\end{equation}
Further identities can be written down in a completely analogous
way~\cite{bjw76,j79}.

%%%%%%%%%%%%%%%%%%%%%%%%%%%%%%%%%%%%%%%%%%%%%%%%%%%%%%%%%%%%%%%%%%%%%%%%%%%%%%%

\section{Two--loop perturbation theory for the two--point vertex functions}
\label{two_loop_vertex_functions}

This appendix comprises the Feynman diagrams to two--loop order for the
(1+1)--dimensional Kardar--Parisi--Zhang equation, and the corresponding
momentum integrals. The integrations over the internal frequencies have already
been performed using the residue theorem.

\begin{figure}
\epsfxsize=\columnwidth\epsfbox{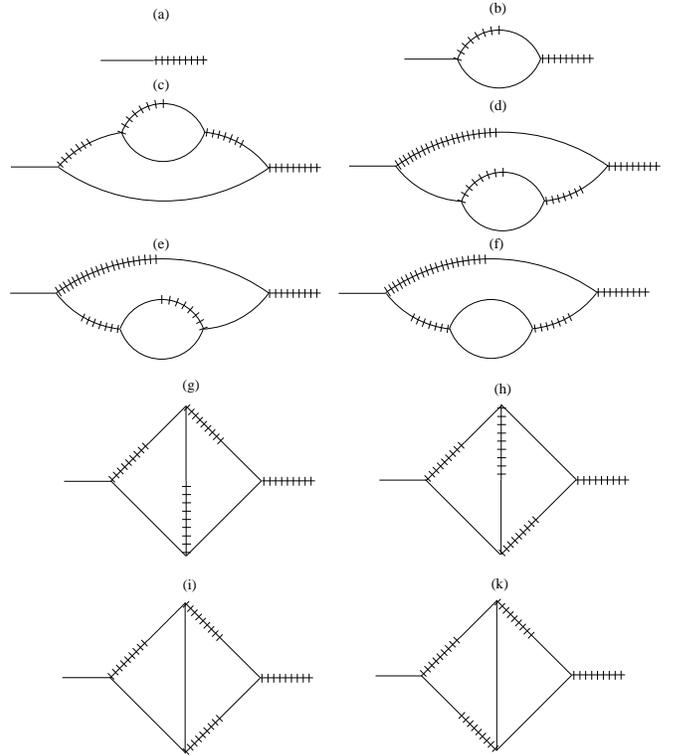}
\vspace{15pt}\caption{Feynman diagrams for the two--point vertex function
  $\Gamma_{{\tilde h} h}(q,\omega)$ to two--loop order.}
\label{2loop_diagrams}
\end{figure}

We start with a list of the contributions to two--loop order to the fully
wavevector-- and frequency--dependent two--point vertex function
$\Gamma_{{\tilde h} h}(q,\omega)$. The other non--vanishing vertex function
$\Gamma_{{\tilde h}{\tilde h}}(q,\omega)$ can be calculated in a similar
fashion (see Ref.~\cite{ft94}), or simply be obtained via the
fluctuation--dissipation theorem (\ref{fdt1}). In writing down the diagrammatic
expansion for the dynamic functional one has to take into account restrictions
which follow from causality. In section II we have not explicitly included the
Jacobian ${\cal J} [h] = {\cal D} [\eta] / {\cal D} [h]$, which depends on the
discretization of the Langevin equation (needed to give a proper definition to
the path integral). As can be shown quite generally~\cite{bjw76}, the Jacobian
cancels the equal--time contractions of the field $h$ and the response field
${\tilde h}$. Keeping this in mind (or by choosing a discretization with the
Jacobian equal to $1$), one can omit the Jacobian in the dynamic functional.
The Feynman diagrams, which account for the restrictions imposed by causality,
are depicted in Fig.~\ref{2loop_diagrams}.
\end{multicols}
\widetext Introducing the abbreviations $q_\pm = (q/2) \pm p$, ${\bar q}_\pm =
(q_-/2) \pm k$, and ${\tilde q}_\pm = q_\pm \pm k$, the corresponding
analytical expressions read:
\begin{eqnarray}
  &&\Gamma_{{\tilde h} h}(q,\omega) \ : \nonumber \\
  &&(a) + (b) = i \omega + \nu_0 q^2 + {\lambda^2 D_0 \over 2 \nu_0} q^2 
    \int_p \, {1 \over i \omega + \nu_0 q_+^2 + \nu_0 q_-^2} \ , \label{A7} \\
  &&(c) = - {\lambda^4 D_0^2 \over 2 \nu_0^2} q
    \int_p \, {q_-^3 \over (i \omega + \nu_0 q_+^2 + \nu_0 q_-^2)^2}
    \int_k \, {1 \over i \omega + \nu_0 q_+^2 +
                  \nu_0 {\bar q}_+^2 + \nu_0 {\bar q}_-^2} \ , \label{A8} \\
  &&(d) = - {\lambda^4 D_0^2 \over 2 \nu_0^2} q
    \int_p \, {q_+ q_-^2 \over i \omega + \nu_0 q_+^2 + \nu_0 q_-^2} \int
    \! dk \, {1 \over \nu_0 q_-^2 + \nu_0 {\bar q}_+^2 + \nu_0 {\bar q}_-^2}
                                                               \label{A9} \\
  &&\phantom{(d) = } \qquad \qquad \times \biggl[ {1 \over 2 \nu_0 q_-^2} +
    {1 \over i \omega + \nu_0 q_+^2 + \nu_0 {\bar q}_+^2 + \nu_0 {\bar q}_-^2}
    \biggl( 1 + {\nu_0 q_-^2 + \nu_0 {\bar q}_+^2 + \nu_0 {\bar q}_-^2 \over
    i \omega + \nu_0 q_+^2 + \nu_0 q_-^2} \biggr) \biggr] \ , \nonumber \\
  &&(e) = - {\lambda^4 D_0^2 \over 2 \nu_0^2} q \int_p \,
    {q_+ q_-^2 \over 2 \nu_0 q_-^2 [i \omega + \nu_0 q_+^2 + \nu_0 q_-^2]} 
    \int_k \, {1 \over \nu_0 q_-^2 + \nu_0 {\bar q}_+^2 + \nu_0 {\bar q}_-^2}
    \ , \label{A10} \\
  &&(f) = {\lambda^4 D_0^2 \over 2 \nu_0^2} q
    \int_p \, {q_+ q_-^2 \over i \omega + \nu_0 q_+^2 + \nu_0 q_-^2}
    \int_k \, {1 \over \nu_0 q_-^2 + \nu_0 {\bar q}_+^2 + \nu_0 {\bar q}_-^2}
    \biggl( {1 \over \nu_0 q_-^2} + {1 \over i \omega + \nu_0 q_+^2 +
    \nu_0 {\bar q}_+^2 + \nu_0 {\bar q}_-^2 } \biggr) \ , \label{A11} \\
  &&(g) = - {\lambda^4 D_0^2 \over \nu_0^2} q
    \int_p \, {q_+ \over i \omega + \nu_0 q_+^2 + \nu_0 q_-^2} \int_k
    \, {{\tilde q}_+ k \over [\nu_0 q_-^2 + \nu_0 {\tilde q}_-^2 + \nu_0 k^2]
    [i \omega + \nu_0 {\tilde q}_+^2 + \nu_0 {\tilde q}_-^2]} \label{A12} \\
  &&\phantom{(g) = } \qquad \qquad \times \biggl( 1 + {2 \nu_0 q_-^2 \over
    i \omega + \nu_0 q_+^2 + \nu_0 {\tilde q}_-^2 + \nu_0 k^2} \biggr) \ ,
    \nonumber \\
  &&(h) = - {\lambda^4 D_0^2 \over \nu_0^2} q
    \int_p \, {q_- \over i \omega + \nu_0 q_+^2 + \nu_0 q_-^2}
    \int_k \, {{\tilde q}_+ k \over 
    [i \omega + \nu_0 {\tilde q}_+^2 + \nu_0 {\tilde q}_-^2] [i \omega +
    \nu_0 q_+^2 + \nu_0 {\tilde q}_-^2 + \nu_0 k^2]} \ , \label{A13} \\
  &&(i) = - {\lambda^4 D_0^2 \over \nu_0^2} q
    \int_p \, {q_- \over i \omega + \nu_0 q_+^2 + \nu_0 q_-^2}
    \int_k \, {{\tilde q}_+ {\tilde q}_- \over
    [i \omega + \nu_0 {\tilde q}_+^2 + \nu_0 {\tilde q}_-^2] [i \omega +
    \nu_0 q_+^2 + \nu_0 {\tilde q}_-^2 + \nu_0 k^2] } \label{A14} \\
  &&\phantom{(i) = } - {\lambda^4 D_0^2 \over \nu_0^2} q
    \int_p \, {q_+ \over i \omega + \nu_0 q_+^2 + \nu_0 q_-^2}
    \int_k \, {{\tilde q}_+ {\tilde q}_- \over
    [\nu_0 q_-^2 + \nu_0 {\tilde q}_-^2 + \nu_0 k^2]
    [i \omega + \nu_0 {\tilde q}_+^2 + \nu_0 {\tilde q}_-^2]} \nonumber \\ 
  &&\phantom{(i) = } \qquad \qquad \times \biggl( 1 + {2 \nu_0 q_-^2 \over
    i \omega + \nu_0 q_+^2 + \nu_0 {\tilde q}_-^2 + \nu_0 k^2} \biggr) \ ,
    \nonumber \\
  &&(j) = {\lambda^4 D_0^2 \over \nu_0^2} q \int_p \, q_+ q_- \int_k \,
    {{\tilde q}_+ \over [\nu_0 q_-^2 + \nu_0 {\tilde q}_-^2 + \nu_0 k^2]
    [i \omega + \nu_0 {\tilde q}_+^2 + \nu_0 {\tilde q}_-^2] [i \omega +
    \nu_0 q_+^2 + \nu_0 {\tilde q}_-^2 + \nu_0 k^2] } \ . \label{A15} 
\end{eqnarray}

\begin{multicols}{2}
  \narrowtext $^*$ Present address: Department of Physics -- Theoretical
  Physics, University of Oxford, 1 Keble Road, Oxford OX1 3NP, U.K.

\end{multicols}
\end{document}